\documentclass[amsmath,amssymb,
reprint
]{revtex4-1}

\usepackage{graphicx}
\usepackage{dcolumn}
\usepackage{bm}

\usepackage[utf8]{inputenc}
\usepackage[T1]{fontenc}
\usepackage{mathptmx}
\usepackage{etoolbox}

\usepackage{xcolor}
\usepackage{lineno,hyperref}

\newcommand{\nn}{\nonumber}
\newcommand{\veps}{\varepsilon}
\newcommand{\tl}{\tilde}

\begin{document}

\title{The effect of ion solvation on ion-induced nucleation -- a generalized Thomson model}

\author{Roni Kroll}
\email{krollr@post.bgu.ac.il.}
\affiliation{Department of Chemical Engineering, Ben-Gurion University of the 
Negev, Beer Sheva 8510501, Israel.}
\author{Yoav Tsori}
\email{tsori@bgu.ac.il.}
\affiliation{Department of Chemical Engineering, Ben-Gurion University of the 
Negev, Beer Sheva 8510501, Israel.}

\date{\today}

\begin{abstract}

We present a model for ion-induced nucleation, focusing on the effect of dissociated ions embedded in 
the fluid surrounding a charged core or colloid. The model includes the ions' direct electrostatic energy 
and preferential solvation. The integrated ions' free energy has two terms: 
The first can be short- or long-range, depending on their density. The second is proportional to the 
nucleus’ volume and can shift the state from undersaturation to supersaturation at high ion 
concentration. The inclusion of the Gibbs transfer energies of ions in the free energy leads to a 
modified Poisson-Boltzmann equation for the potential around the core. The integrated ions' free 
energy is added to the fluids' interfacial and bulk terms to establish a generalized Thomson model. 
In the Debye-H\"{u}ckel limit, the model is solved analytically, while in the nonlinear regime, it is solved 
numerically. The state diagram in the plane of saturation and core charge includes regions with a 
homogeneous phase, electro-prewetting, metastable vapor, metastable nucleus, and spontaneous 
nucleation states. The lines separating these regions depend sensitively on the preferential solvation.
Our model shows nucleation asymmetry to the sign of the nucleus' charge. This sign 
asymmetry is due to the Gibbs transfer energies of ions.

\end{abstract}


\maketitle


\section{Introduction}

Nucleation is a process where a stable phase emerges from a metastable state. It occurs in 
crystallization, melting, and boiling in physical, chemical and biological systems.\cite{kashchiev_book}
Nucleation depends on the microscopic, molecular, details, and therefore nucleation rates are difficult 
to obtain.\cite{michaelides_chem_rev_2016} Classical nucleation theory is a continuum approach that 
uses the capillary assumption to describe the incumbent mesoscopic nucleus and the metastable 
phase by their bulk properties.\cite{volmer_cnt_1926,becker_cnt_1935,frenkel_cnt_1939} The theory 
assumes an infinitely sharp interface between phases and expresses the phenomenological free 
energy curve as a function of nucleus size $R$.\cite{kalaichelvi_review_2016,abraham_book} 
Heterogeneous nucleation around small colloidal particles, e.g., aerosols in the 
atmosphere,\cite{curtius_gra_2004,kenjiro_gra_2006,kreiden_atm_chem_phys_2007,kreiden_chem_rev_2015}
typically 
reduces the energy barrier between the phases and increases the corresponding transition 
rates.\cite{oxtoby_anpc_1995,mahiddine_chem_rev_2014} 
When a charged particle is embedded in vapor, a dielectrophoretic force acts on the fluid, aiding liquid 
condensation near the particle. The 
force and electrostatic potential decay as the inverse distance from the particle. Thomson's model 
incorporates this effect by adding a term proportional to the particle's charge squared into the classical 
nucleation theory.\cite{Thomson_1933,podguzova_atm_res_2011} Indeed, experiments show 
that ion-induced nucleation, taking place in the presence of charged molecular clusters, is significantly 
enhanced. \cite{Henrik_2011,ADACHI_1992,Curtius_2006}

In a recent work, we extended the Thomson model to describe nucleation around charged 
particles in fluids containing dissolved ions, considering only the electrostatic forces.\cite{KROLL_2023} 
In such fluids, dissociated ions lead to 
screening of the potential characterized by the Debye length. We found that the existence of ions 
lowers the energy barrier for nucleation. 
In the limit of high ionic strengths, the field is screened over a short length scale, and the effect of the 
charged particle is proportional to its {\it surface}. 
At negligible salt content, the behavior becomes identical to the Thomson model.

The term preferential ion solvation relates to the different ability of ions to dissolve in different 
solvents. Ion-specific effects are evident in the Hofmeister series and are 
well-studied.\cite{Hofmeister_1888,ZHANG_2006,Gregory_2021} Ammonium, sulfuric acid, and highly 
oxidized biogenic vapors are among the studied ions.\cite{rabeony_1987,Almeida_2013,Kirby_2016} 
In liquid mixtures, the preferential solvation of ions depends on several parameters, such as the type 
of noncovalent bonds, ion size, and polarity.\cite{Marcus_1989,Chan_1994} 
Preferential solvation affects several properties, such as surface tension and phase 
transitions.\cite{Onuki_2008,Onuki_2006} The ionic preference is quantified by the Gibbs transfer 
energy required for an ion to move from one phase to another. 

The influence of charge sign on ion-induced nucleation has been extensively studied both 
experimentally and theoretically. Wilson's pioneering work demonstrated that nucleation is more 
pronounced around negative ions than positive ones.\cite{Wilson_1897}
For organic compounds such as dibutylpthalate, cations were found to promote nucleation more 
than anions, whereas the opposite trend was observed for nucleation of \textit{n}-propanol around 
inorganic seed aerosol.\cite{ADACHI_1992,winkler_2008}
Recent research has shown no clear sign preference for sub-3 nm particles and suggests that the 
parameter important in this size range is not the electrical sign but the chemical 
composition.\cite{Kangasluoma_2016}
Several theoretical studies have attempted to elucidate the origin of sign symmetry-breaking using 
Monte Carlo and molecular dynamics simulations and density functional theory, but the conclusions 
were not definitive. \cite{Oh_2001,Laaksonen_2002,Nadykto_2008,kathmann_2005,chen_jcp_2012}

Here, we generalize our previous work and look at the importance of ion solvation to nucleation. 
The free energy of the ions is composed of their entropy, electrostatic interactions, and preferential 
solvation. Ionic solvation appears as a {\it volume} term, and thus its effect occurs far from 
the particle. The ions' contribution is added to the classical surface tension and bulk terms in the 
energy to yield the Gibbs energy difference for nucleation. We show that core charge-asymmetric 
nucleation can be the result of an asymmetry between the Gibbs solvation 
energies of cations and anions.

\section{Model description}

\begin{figure}[h!]
\centering
\includegraphics[width=0.49\textwidth,clip]{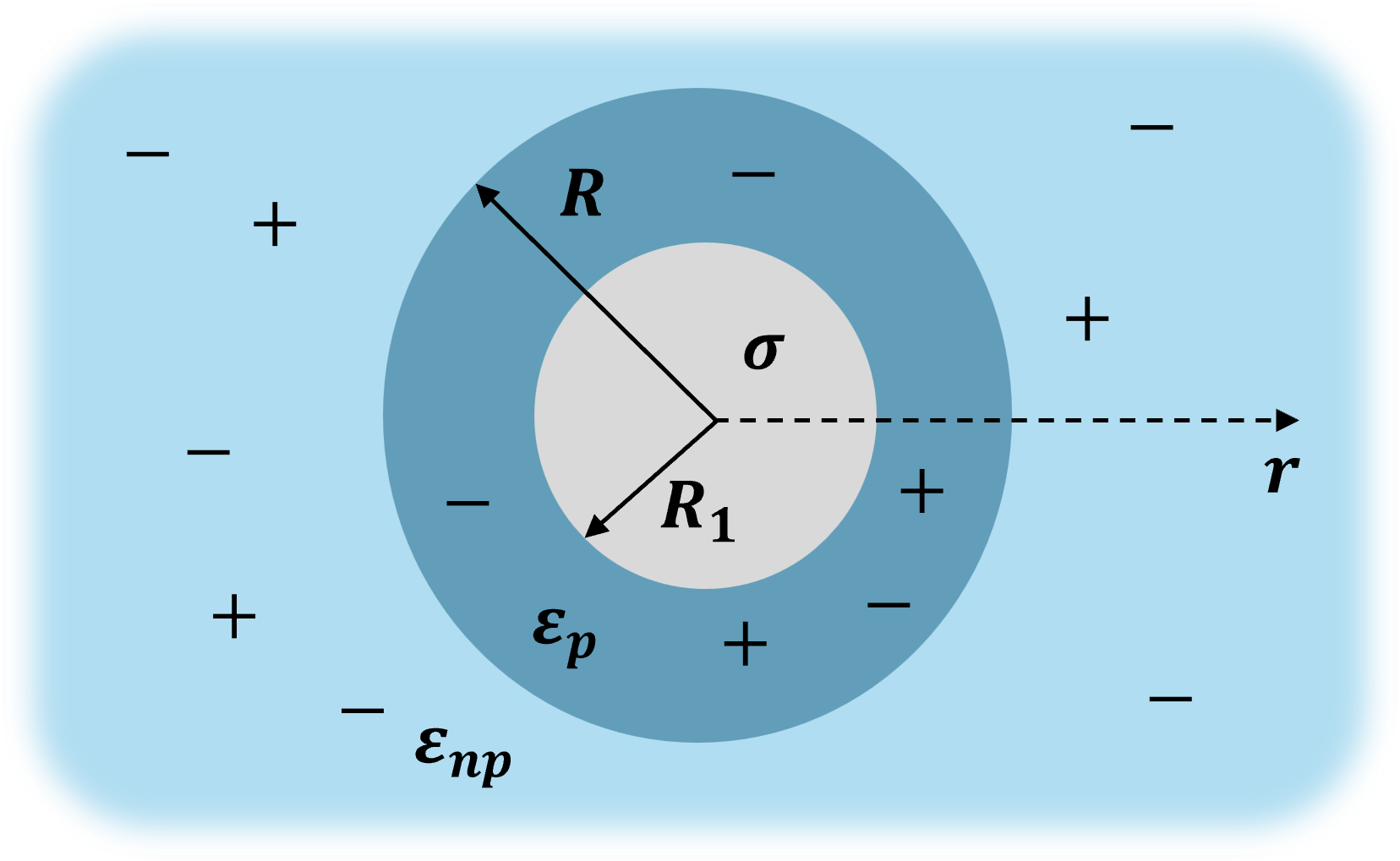}
\caption{Model illustration. A charged core particle (gray) with radius $R_1$ and surface charge density 
$\sigma$, surrounded by a polar phase (dark blue) with radius $R$ and a nonpolar phase (light blue). 
The interface between the phases is infinitely thin. The system contains ions in both phases.}
\label{model}
\end{figure}

Figure \ref{model} illustrates a spherical charged particle (colloid) with a uniform surface charge 
density $\sigma$ and radius $R_1$, surrounded by two immiscible fluids. Due to dielectrophoretic 
forces, the polar fluid is in contact with the colloid, while farther away the nonpolar phase is 
present. The system is spherically symmetric and the sharp interface between the two fluids is at 
$r=R$, where $r$ is the distance from the core's center. The relative permittivities of the polar and 
nonpolar fluids are $\veps_{\rm p}$ and $\veps_{\rm np}$. Ions are present in both phases but have a 
larger affinity to the polar phase. The preferential solvation of the ions is quantified by 
the Gibbs transfer energy, reflecting the difference in the equilibrium ions solvation energies 
between two bulk solvents. In our study, it is defined as the solvation energy of an ion in the polar 
phase minus its solvation energy in the nonpolar phase.
For example, the Gibbs transfer energy of a sodium ion from water to \textit{n}-propanol is $\Delta 
G_t=16.8 [kJ/mol]$ and for chloride ion it is $\Delta G_t=25.5 [kJ/mol]$.\cite{Marcus_1980}

Note that even without a charged particle, close to the critical 
point, the presence of ions ``help'' condensation of gas to liquid; they raise the critical temperature by 
an amount proportional to $\Delta G_t^2$.\cite{Tsori_2007_2} In this work, we assume that the density 
and temperature are far from the critical point and therefore ignore this effect.

In our generalized Thomson model, the change in the Gibbs free energy for the formation of a dense, 
polar layer, at $r=R$ is
\begin{eqnarray}
\Delta G=4\pi\gamma R^2-\frac{4}{3}\pi\left(R^3-R_1^3\right)\Delta\mu+F_{\rm ions}. 
\label{Gibbs}
\end{eqnarray}
The first and second terms are the surface tension and bulk energy terms, respectively, where 
$\gamma$ is the surface tension between the two phases, and $\Delta \mu$ is the difference in the
chemical potential difference of the fluids per unit volume. 
When $\Delta \mu$ is positive, the mixture is supersaturated; when it is negative, it is undersaturated. 
The third term is the added energy due to the presence of the ions and the charged core. 
The dissolved ions are assumed to be point-like and monovalent. We write $F_{\rm ions}$ as a volume 
integral of the entropy, solvation, and electrostatic energy densities as
\begin{align}
F_{\rm ions}=\int \biggl\{k_BT\left[n^+(\ln(v_0n^+)-1)+n^-(\ln(v_0n^-)-1)\right.\nn\\
+\left.(H(r-R)-1/2)(\Delta u^+n^++\Delta u^-n^-)\right]\nn\\
+\frac12\veps\veps_0(\nabla\psi)^2-\mu^+n^+-\mu^-n^- \biggr\}d{\bf r},
\label{free_energy_int}
\end{align}
where $k_B$ is the Boltzmann constant, $T$ is the absolute temperature, $v_0$ is a 
volume comparable to the ion size, and $H$ is the Heaviside step function. $\psi$ represents the 
electrostatic potential while $n^\pm$ denote the density profiles of anions and cations, all of which are 
dependent on $r$. $\mu^\pm$ are the chemical potentials of the ions, $\veps_0$ is the 
permittivity of the vacuum and $\veps$ is the relative dielectric constant.
The dimensionless parameters $\Delta u^\pm$ quantify the Gibbs transfer energies. For the sodium 
and chloride ions in n-propanol, they are $\Delta u^+\approx 7$ and $\Delta 
u^-\approx 10$, respectively. 

In the regular Thomson model (no ions in the fluids),  
$F_{\rm ions}$ reduces to the integral over the electrostatic energy density only. Here, 
minimization of the free energy with respect to the cation and anion densities yields the Boltzmann 
distributions 
\begin{eqnarray}\label{Density}
n^\pm=n_0 e^{-\Delta u^\pm(H(r-R)-1)\mp\frac{e\psi}{k_BT}},
\end{eqnarray}
with $n_0$ being the ion density at $r\to\infty$ where $\psi=0$.
We start by setting the Gibbs transfer energies equal for the anion and cation, $\Delta u^+=\Delta 
u^-=\Delta u$, but relax this assumption later. 

In spherical symmetry, the electrostatic potential satisfies the modified Poisson-Boltzmann equation
\begin{eqnarray}\label{PB}
\psi_i''+\frac{2}{r}\psi_i'=\frac{2en_0e^{-\Delta 
u(H(r-R)-1)}}{\veps_i\veps_0}\sinh\left(\frac{e\psi}{k_BT}\right),
\end{eqnarray}
where $i$=p or $i$=np for either the polar or nonpolar phases. 

The boundary conditions are as follows: Gauss's law determines the electric field on the colloid's surface, $\psi'_{\rm 
p}(r=R_1)=-\sigma/\veps_0\veps_{\rm p}$. At the interface, the electrostatic potential and 
displacement field are continuous: $\psi_{\rm p}(r=R)=\psi_{\rm np}(r=R)$ and $\veps_{\rm p}\psi_{\rm 
p}'(r=R)=\veps_{\rm np}\psi_{\rm np}'(r=R)$, and far away from the charged colloid the potential decays to zero, $\psi_{\rm np}(r\rightarrow \infty)=0$. 

When the ion densities in Eq. (\ref{Density}) are substituted in Eq. (\ref{free_energy_int}), the 
dimensionless energy of the ions can be written as
\begin{eqnarray}
\frac{F_{\rm ions}}{4\pi k_BT}&=&\tl{n}_0\int\left\{2e^{-\Delta 
u(H(\tl{r}-\tl{R})-1)}(\tl{\psi}\sinh(\tl{\psi})-\cosh(\tl{\psi}))\right.\nn\\
&+&\left.\veps_i\tl{\lambda}_0^2(\tl{\nabla}\tl{\psi})^2
\right\}\tl{r}^2d{\tl{r}},
\label{free_energy_short}
\end{eqnarray}
where $\tl{\psi}=e\psi/k_BT$ and $\tl{n}_0=n_0R_1^3$. 
The ``vacuum Debye length'' is defined with the permittivity of the vacuum as 
$\lambda_0=\left(\veps_0k_BT/2n_0e^2\right)^{1/2}$, and all lengths with ''$\sim$'' sign are scaled by 
$R_1$.

\section{Debye-H\"{u}ckel limit}

In the Debye-H\"{u}ckel limit, $\tl{\psi}\ll 1$, the Poisson-Boltzmann equation (\ref{PB}) can be 
linearized to yield 
\begin{equation}
\tl{\psi}_i''+\frac{2}{\tl{r}}\tl{\psi}_i'-\frac{1}{\tl{\lambda}_i^2}\tl{\psi}_i=0,
\label{DH}
\end{equation}
where the Debye lengths in the polar and nonpolar phases are defined by
\begin{eqnarray}
\lambda_p^2=\frac{\veps_p}{e^{\Delta u}}\lambda_0^2,~~~~
\lambda_{\rm np}^2=\veps_{\rm np}\lambda_0^2, 
\end{eqnarray}
and their scaled versions are $\tl{\lambda}_i=\lambda_i/R_1$. Note that the exponential dependence of 
$\lambda_p$ on $\Delta u$ can reduce it significantly.
The analytical solution is a combination of Yukawa potentials\cite{KROLL_2023}
\begin{eqnarray}\label{psi_two_regions}
\tl{\psi}_{\rm p}&=&\tl{\sigma}\tl{a}_{\rm p}\frac{e^{\tl{r}/\tl{\lambda}_{\rm 
p}}}{\tl{r}}+\tl{\sigma}\tl{b}_{\rm 
p}\frac{e^{-\tl{r}/\tl{\lambda}_{\rm p}}}{\tl{r}},~~~~~~~~1\leq
\tl{r}\leq \tl{R},\nn\\
\tl{\psi}_{\rm np}&=&\tl{\sigma}\tl{b}_{\rm np}\frac{e^{-\tl{r}/\tl{\lambda}_{\rm 
np}}}{\tl{r}},~~~~~~~~~~~~~~~~~~~~~~~~~~\tl{r}\geq \tl{R},
\end{eqnarray}
where $\tl{\sigma}=\sigma e R_1/(k_BT\veps_0)$. The boundary condition at 
$\tl{r}\rightarrow\infty$ eliminates the diverging exponential in $\tl{\psi}_{\rm np}$. 
The boundary conditions lead to a system of three linear equations with three unknown 
parameters, $\tl{a}_{\rm p}$, $\tl{b}_{\rm p}$, and $\tl{b}_{\rm np}$, that can readily be solved.

Once these coefficients are determined, the energy $F_{\rm ions}$ in Eq. 
(\ref{free_energy_int}) can be written analytically by
\begin{eqnarray}
\frac{F_{\rm ions}}{4\pi k_BT}&=&-\frac23\tl{n}_0\left(\tl{R}^3-1\right)(e^{\Delta 
u}-1)+p_{\sigma}h(\tl{R}).
\end{eqnarray}
The first term is the volume contribution of the solvation energy. It expresses a driving 
force that prefers a polar phase with condensed ions. It is linear in $\Delta u$ for small values of 
$\Delta u$. The second term is the electrostatic energy, where $p_{\sigma}=\sigma^2 
R_1^3/2k_BT\veps_0$ is the dimensionless ratio between the electrostatic energy stored in a sphere of 
radius $R_1$ and the thermal energy. The value of $p_{\sigma}$ can significantly vary; for example, 
when $\sigma$ corresponds to $8$ unit charges evenly distributed over a surface of a sphere with 
$R_1=2$ nm, at room temperature $p_{\sigma}\approx 70$, whereas for a radius of $R_1=50$ nm, 
$p_{\sigma}\approx 3$.

$h(\tl{R})$ above is a function of the nucleus size given by\cite{KROLL_2023}
\begin{eqnarray}
h(\tl{R})&=&\veps_p\left[\tl{a}_{\rm p}^2e^{2\tl{r}/\tl{\lambda}_{\rm p}}(1/\tl{\lambda}_{\rm 
p}-1/\tl{r})-\tl{b}_{\rm p}^2e^{
-2\tl{r}/\tl{\lambda}_{\rm p}}(1/\tl{\lambda}_{\rm p}+1/\tl{r})\right.\nn\\
&-&\left.2\tl{a}_{\rm p}\tl{b}_{\rm 
p}/\tl{r}\right]_{\tl{r}=1}^{\tl{r}=\tl{R}}
+\veps_{\rm np}\tl{b}_{\rm np}^2e^{-2\tl{R}/\tl{\lambda}_{\rm np}}(1/\tl{\lambda}_{\rm 
np}+1/\tl{R}).
\end{eqnarray}
This Debye-H\"{u}ckel expression is valid for small surface charges. Its dependence on $\tl{R}$ can be 
strong or weak. 

The total Gibbs nucleation energy Eq. (\ref{Gibbs}) can now be written in dimensionless form as
\begin{eqnarray}
\frac{\Delta G}{4\pi k_BT}&=&\\
p_{\gamma}\tl{R}^2&-&\left(\tl{R}^3-1\right)\left(p_{\mu}+\frac23 
\tl{n}_0(e^{\Delta u}-1)\right)+p_{\sigma}h(\tl{R}),\nn
\label{Full_Gibbs}
\end{eqnarray}
where $p_{\gamma}=\gamma R_1^2/k_BT$ and $p_{\mu}=\Delta \mu R_1^3/3k_BT$. 

At the limit of high salt concentration, $\tl{\lambda}_{\rm p}\ll 1$, the electric field is screened 
over the short distance $\tl{\lambda}_p$, and $h(\tl{R})$ tends to $h(\tl{R})\approx \tl{\lambda}_{\rm 
p}/\veps_{\rm p}$ for $\tl{R}>1+\tl{\lambda}_p$. In this limit, the $p_{\sigma}h(\tl{R})$ 
term in Eq. (\ref{Full_Gibbs}) is independent of $\tl{R}$ and plays the role of a surface term, i.e., the 
energy $\Delta G$ is an {\it effective} surface tension:
\begin{eqnarray}
\frac{\Delta G}{4\pi R_1^2}=\gamma_{\rm ls}+\frac{\sigma^2\lambda_{\rm p}}{2\veps_0\veps_{\rm p}},
\label{eff_ST}
\end{eqnarray}
where we added the bare surface tension between the liquid and the solid colloid $\gamma_{\rm ls}$.  
The correction to the bare surface tension scales with $\Delta u$ as 
$\sim\lambda_{\rm p}\propto e^{-\Delta u/2}$. 

In the opposite limit of low salt concentration, $\tl{\lambda}_i\to \infty$, the electrostatic term becomes
\begin{eqnarray}
p_{\sigma}h(\tl{R})\approx p_{\sigma}\left[-\frac{1}{\veps_{\rm 
p}}\left(\frac{1}{\tl{R}}-1\right)+\frac{1}{\veps_{\rm np}\tl{R}}\right],
\end{eqnarray}
which is identical to the Thomson expression for dielectric fluids.

We note that the volume term in Eq. (\ref{Full_Gibbs}) ($\propto \tl{R}^3-1$) is strong and long-range at 
high salt content; it leads to a renormalization of the supersaturation $p_\mu$. Indeed, by combining 
the $p_\mu$ and $\sim\tl{n}_0$ terms, one can write an {\it effective} chemical potential difference as
\begin{eqnarray}
\mu_{\rm eff}=\mu+2k_BT n_0\left(e^{\Delta u}-1\right).
\label{eff_mu}
\end{eqnarray}
The second term on the right-hand side is of order $k_BT$ and can be large due to the exponential 
dependence on $\Delta u$.

In the next section, we solve the nonlinear problem numerically and obtain the energy for higher 
values of $\sigma$.

\section{Numerical results}

When the Debye-H\"{u}ckel limit does not hold, we solve Eq. (\ref{PB}) numerically, with the charge 
density in the boundary condition at $\tl{r}=1$ determined by $p_{\sigma}$.
We set values for the Debye length in vacuum, $\lambda_0$, and the  Bjerrum length in a vacuum, 
$l_{B0}=e^2/\veps_0 k_BT$, which is the distance at which the thermal energy equals the Coulombic 
interaction of two unit charges. This indirectly determines the bulk ion density $n_0$ via 
$n_0=1/2l_{B0}\lambda_0^2$. 

For the results in this section, we used $l_{B0}/R_1=70$, $\lambda_0/R_1=5$, and core with radius 
$R_1\approx 10$ nm, and therefore the ion concentration is $n_0\approx5\times10^{-9}$ M. For the 
values $\veps_p=80$ and $\Delta u=2$ (see figures below), we find that $\lambda_p>l_{B0}$. Ion-ion 
correlations are thus small and the use of the mean-field Poisson-Boltzmann approach is justified.

\begin{figure}[h!]
\centering
\includegraphics[width=0.49\textwidth,bb=90 230 490 540,clip]{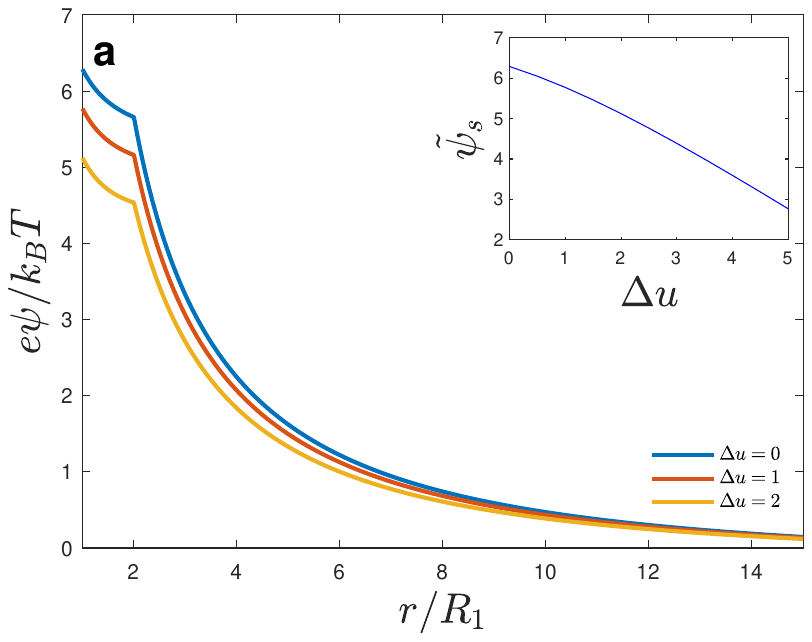}
\includegraphics[width=0.49\textwidth,bb=90 240 488 540,clip]{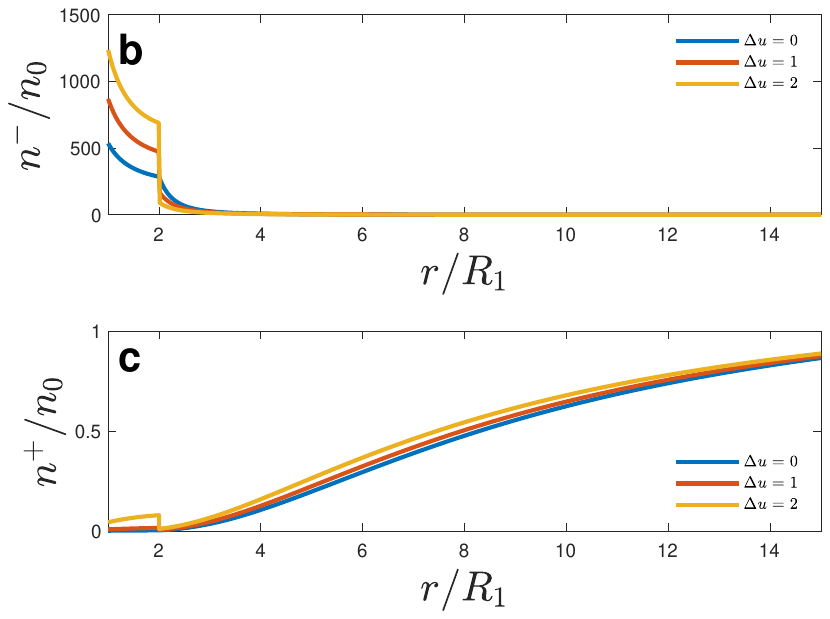}
\caption{(a) Dimensionless electrostatic potential profiles for different values of $\Delta u$, calculated 
numerically from Eq. (\ref{PB}). The inset shows the dimensionless surface potential 
$\tl{\psi}_s=e\psi(r=R_1)/k_BT$, as a function of $\Delta u$. 
(b) Anion and (c) cation density profiles from Eq. (\ref{Density}). In (b), the high anion density at small 
$r/R_1$ is due to the assumption of point-like ions. In all parts of the figure, 
the nucleus' radius is $R=2R_1$, where $R_1$ is the core radius, $p_{\sigma}=80$, $\lambda_0/R_1=5$, 
$l_{B0}/R_1=70$, $\veps_{\rm p}=80$ and $\veps_{\rm np}=6$.}
\label{pot_sol}
\end{figure}

Potential profiles for a given nucleus of size $R=2R_1$ are presented in Fig. \ref{pot_sol}(a).
The profiles $\psi(r)$ are continuous; the electric field $E=-\psi'(r)$ is discontinuous across the interface 
due to the jump in permittivity. The inset shows that with increasing values of $\Delta u$ (increase in 
the  affinity of ions to the polar phase at $r<R$) the surface potential, $\psi_s$, decreases. 
The corresponding ion distributions are presented in Figs. \ref{pot_sol}(b) and \ref{pot_sol}(c).
The densities are continuous across the interface in the absence of preferential solvation ($\Delta 
u=0$). However, when the ions are preferentially miscible in the polar phase $\Delta u>0$, their 
density in the polar phase increases, accompanied by a jump occurring at the interface. 
This kink results from an effectively chemical potential difference between the phases. 
The interface separates unequal quantities of the ions in the fluids, similar to the Donnan 
potential.\cite{Donnan_1924}
The kink disappears in models where the density of the fluid varies smoothly across the 
interface.\cite{Onuki_2006} 
Although the value of preferential solvation of the anions and cations is similar, 
the difference between their density in the condensed phase is due to the positive sign of the core 
charge. 

To obtain the ions' energy $F_{\rm ions}$ as a function of interface location $\tl{R}$, we 
substitute the potential profiles for different values of $\tl{R}$ into Eq. (\ref{free_energy_short}).
We subtract the energy of a charged sphere surrounded by the nonpolar phase only, $F_{\rm 
ions}(\tl{R}=1)$. 
\begin{figure*}[ht!]
\includegraphics[width=0.49\textwidth,bb=90 230 500 540,clip]{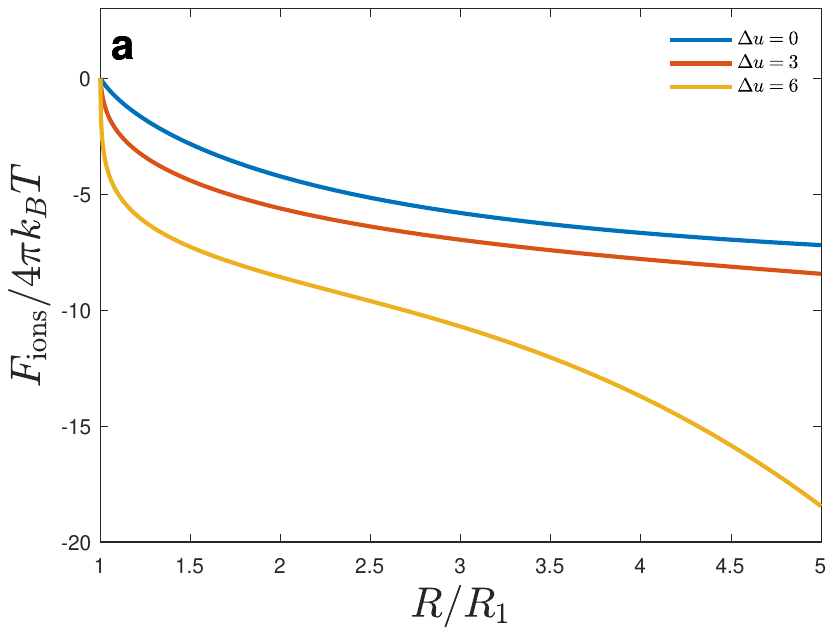}
\includegraphics[width=0.49\textwidth,bb=90 230 500 540,clip]{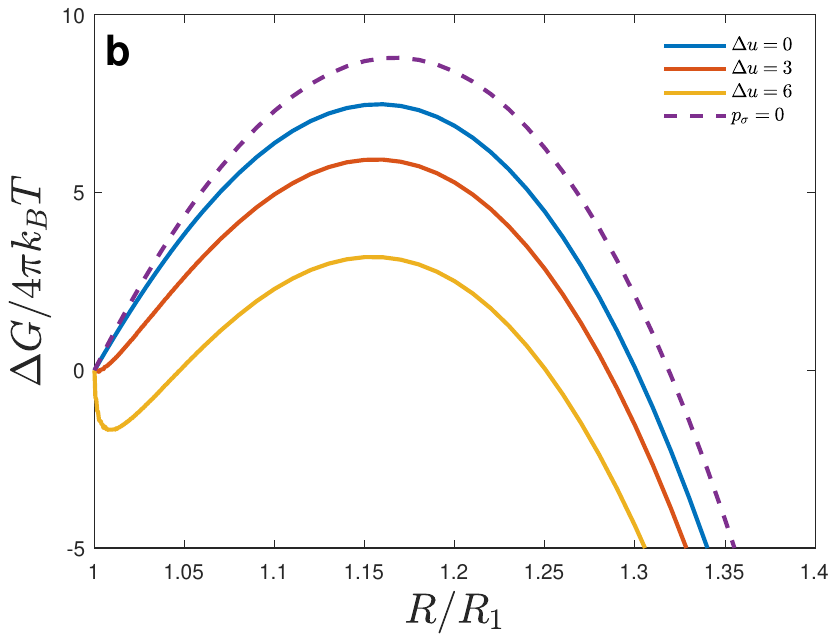}
\includegraphics[width=0.49\textwidth,bb=90 230 500 540,clip]{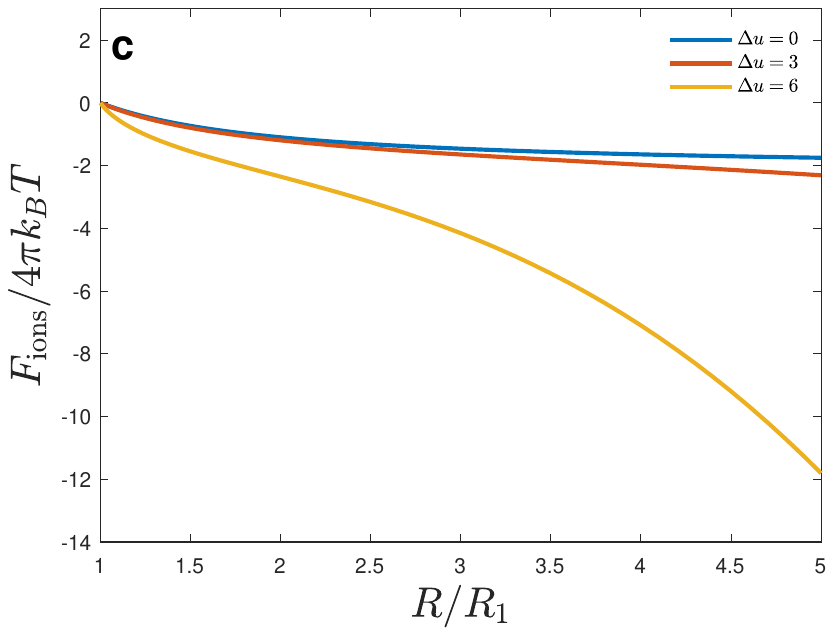}
\includegraphics[width=0.49\textwidth,bb=90 230 500 540,clip]{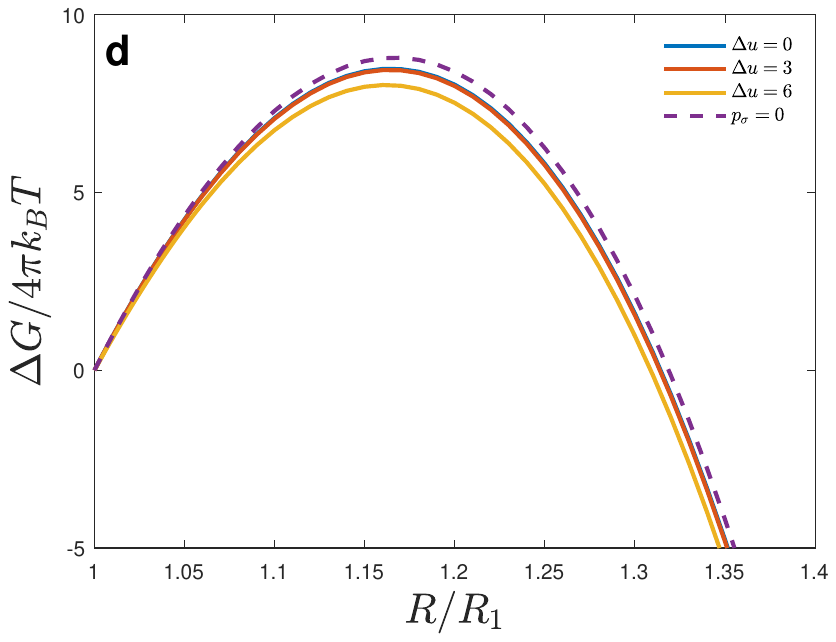}
\caption{The free energy of the ions $F_{\rm ions}/4\pi k_BT$ (a) and the Gibbs nucleation energy 
$\Delta G/4\pi k_BT$ (b), obtained numerically, against the scaled nucleus radius $R/R_1$ ($R_1$ is the 
core radius), with $\veps_{\rm p}=80$ and $\veps_{\rm np}=6$. Figures (c) and (d) are similar, but 
calculated for a higher permittivity of the nonpolar phase $\veps_{\rm np}=25$. We used 
$p_{\sigma}=80$, $p_{\gamma}=350$, $p_{\mu}=200$, $\lambda_0/R_1=5$, and $l_{B0}/R_1=70$.} 
\label{Energies}
\end{figure*}
The curves are presented in Fig. \ref{Energies}(a) for different $\Delta u$ values. 
When $\Delta u=0$, $F_{\rm ions}(\tl{R})$ monotonically decreases with $\tl{R}$. 
When $\Delta u$ increases, the ions' preferential solvation further reduces the energy, 
promoting the growth of the polar phase. 
An inflection point in the energy curves appears due to the large contribution of the volume term to 
the energy from a certain nucleus size (yellow line).

In Fig. \ref{Energies}(b), the total Gibbs energy profiles from Eqs. (\ref{Gibbs}) and 
(\ref{free_energy_short}), including the surface tension 
and bulk energies, are shown. The dashed line is the energy curve for classical nucleation when the 
colloid is uncharged, $p_{\sigma}=0$. Clearly, the presence of a charged colloid, $p_{\sigma}>0$, 
significantly reduces the nucleation energy barrier. 
Additionally, a slight decrease in the size of the critical radius is observed. 
A local minimum appears in 
$\Delta G$ at small values of $\tl{R}$. This corresponds to a metastable state of a nucleus with a finite 
size. When $\Delta u$ is increased, the size of the metastable nucleus grows, and its energy 
decreases. Figures. \ref{Energies}(c) and \ref{Energies}(d) are similar to parts a-b, but with a higher 
value of $\veps_{\rm np}$. 
In this case, fewer ions are accumulated in the polar phase and the ions' energy, while still 
decreasing with $\tl{R}$, does so more moderately. Consequently, the decrease in the energy barrier is 
smaller, and there are no metastable states in these parameter values.  
\begin{figure}[ht!]
\centering
\includegraphics[width=0.49\textwidth,bb=90 230 500 540,clip]{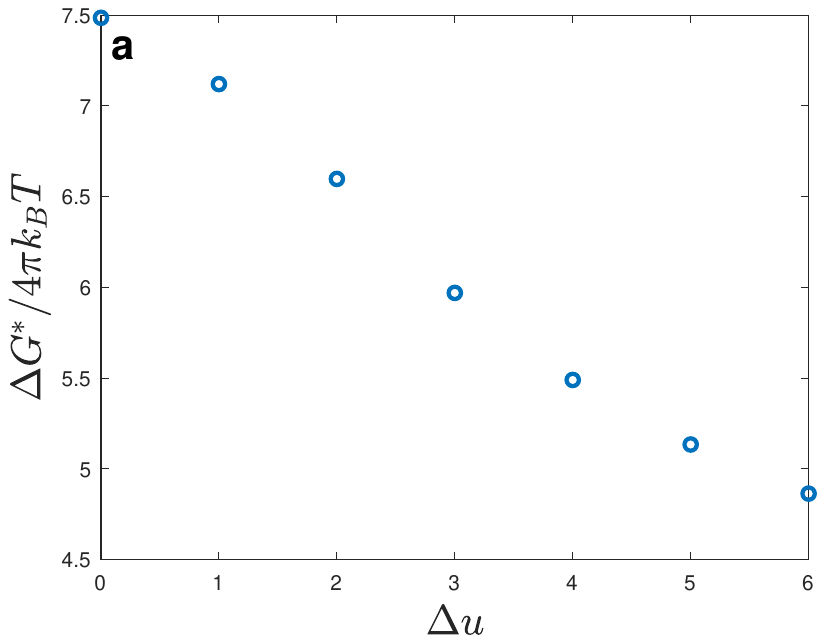}
\includegraphics[width=0.49\textwidth,bb=90 230 500 540,clip]{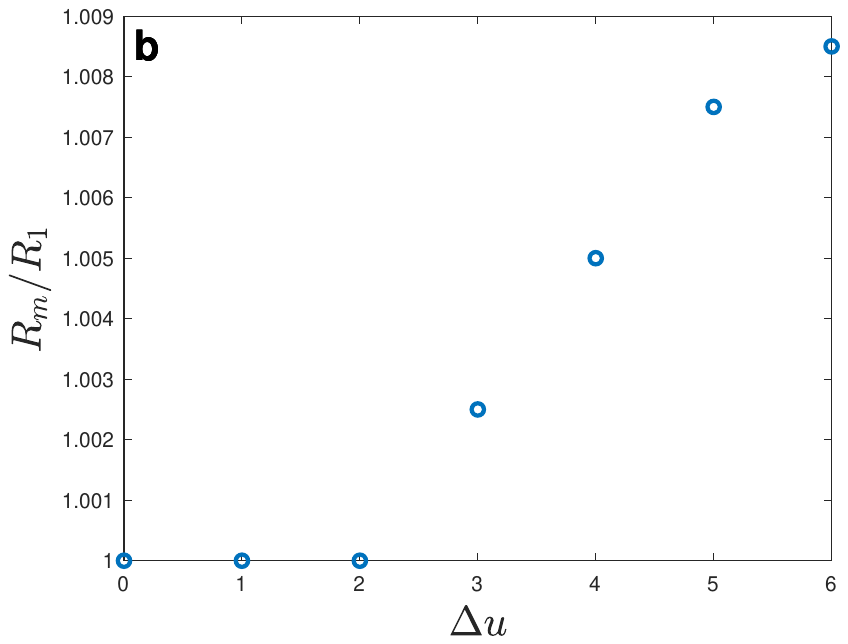}
\caption{(a) The dimensionless nucleation barrier $\Delta G^*/4\pi k_BT$ and (b) the radius of the 
metastable nucleus, $R_m/R_1$ vs Gibbs transfer energy value $\Delta u$. We used $p_{\sigma}=80$, 
$p_{\gamma}=350$, $p_{\mu}=200$, $\lambda_0/R_1=5$, $l_{B0}/R_1=70$, $\veps_{\rm p}=80$ and 
$\veps_{\rm np}=6$.}
\label{Energy_barrier}
\end{figure}

The energy barrier, $\Delta G^*$, is the difference between the maximum and local 
minimum values in the $\Delta G$ curves. The reduction of $\Delta G^*$ as a function of increasing 
$\Delta u$ is presented in Fig. \ref{Energy_barrier}(a). This reduction can significantly accelerate 
nucleation, recalling that the nucleation rate depends exponentially on $\Delta G^*$: $\sim\exp(-\Delta 
G^*/k_BT)$. Figure \ref{Energy_barrier}(b) shows the size of the metastable radius, $R_m$, as a 
function of $\Delta u$. Although the values of the metastable nuclei radius sizes presented here are 
minimal, they can be much larger closer to the critical temperature of the fluids.\cite{tsori_jcp_2021}

\begin{figure}[ht!]
\centering
\includegraphics[width=0.49\textwidth,bb=90 230 500 540,clip]{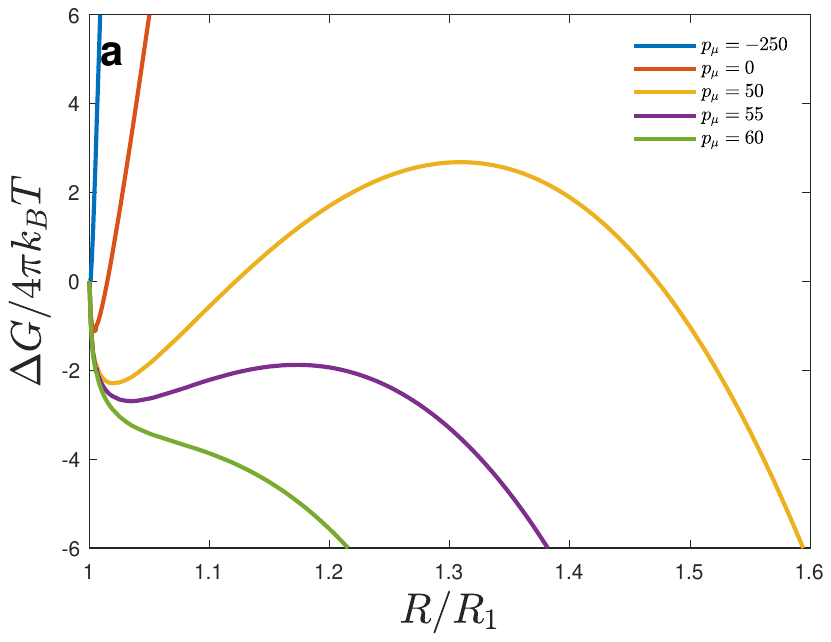}
\includegraphics[width=0.49\textwidth,bb=90 230 500 540,clip]{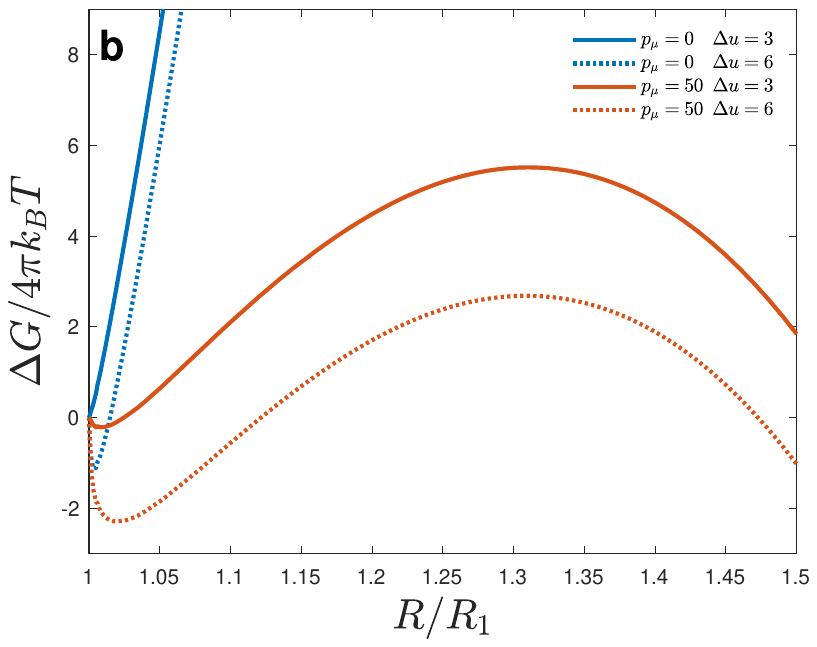}
\caption{(a) Gibbs energy profiles for different values 
of $p_{\mu}$ and $\Delta u=6$. (b) Comparison between Gibbs energy profiles with two values of 
$\Delta u$ and two values of $p_\mu$. We used $p_{\sigma}=80$, $p_{\gamma}=100$, 
$\lambda_0/R_1=5$, $l_{B0}/R_1=70$, $\veps_{\rm p}=80$ and $\veps_{\rm np}=6$.}
\label{Saturation_change}
\end{figure}

We now turn to investigate the effect of fluid saturation, as given by the parameter $p_{\mu}$.
When $p_{\mu}<0$, the fluid is undersaturated and surrounded only by a nonpolar phase.
$p_{\mu}=0$ is the coexistence (binodal) value, and when $p_{\mu}>0$, the fluid is supersaturated. 
Figure \ref{Saturation_change}(a) shows the Gibbs nucleation curves for varying supersaturation 
values. 
$\Delta G$ for an undersaturated term is represented by the blue line, where the nucleation energy 
increases with the nucleus size due to the positive surface tension and bulk terms. Under these 
conditions, the electrostatic energy is 
insufficient to create a nucleus. For a positive value of $p_{\mu}$, a local minimum appears at small 
$\tl{R}$ values and a maximum at larger $\tl{R}$s. An increase in $p_{\mu}$ reduces the energy barrier 
until, at some point, it disappears. This is the value of the ``electrostatic spinodal'', defined by 
$\Delta G^*=0$, or, equivalently, by the existence of $\tl{R}$ for which 
$\Delta\tl{G}''(\tl{R})=\Delta\tl{G}'(\tl{R})=0$. The electrostatic spinodal is a generalization of the regular 
spinodal curve for spontaneous nucleation due to the presence of the charged core and ions. Figure 
\ref{Saturation_change}(b) compares $\Delta G$ curves for two different saturation values and two 
values of $\Delta u$. When $p_\mu=0$, the blue curves indicate that an increase in $\Delta u$ leads to 
the appearance of a finite-sized polar nucleus. The orange lines describe a supersaturated fluid, 
$p_\mu=50$, where the increase in $\Delta u$ shifts the radius of the metastable nucleus to larger 
values and decreases the nucleation barrier.

\begin{figure}[ht!]
\centering
\includegraphics[width=0.49\textwidth,bb=85 230 500 541,clip]{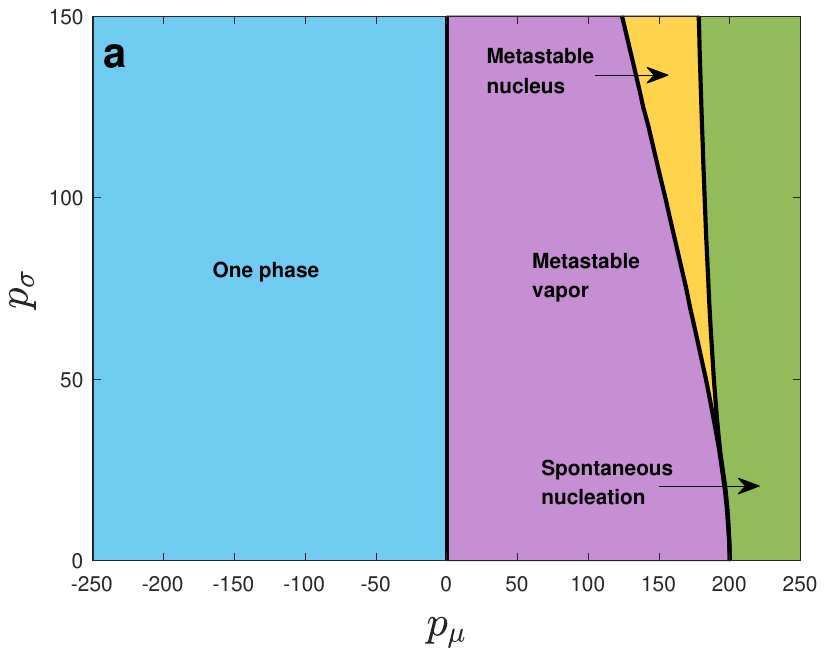}
\includegraphics[width=0.49\textwidth,bb=85 230 500 541,clip]{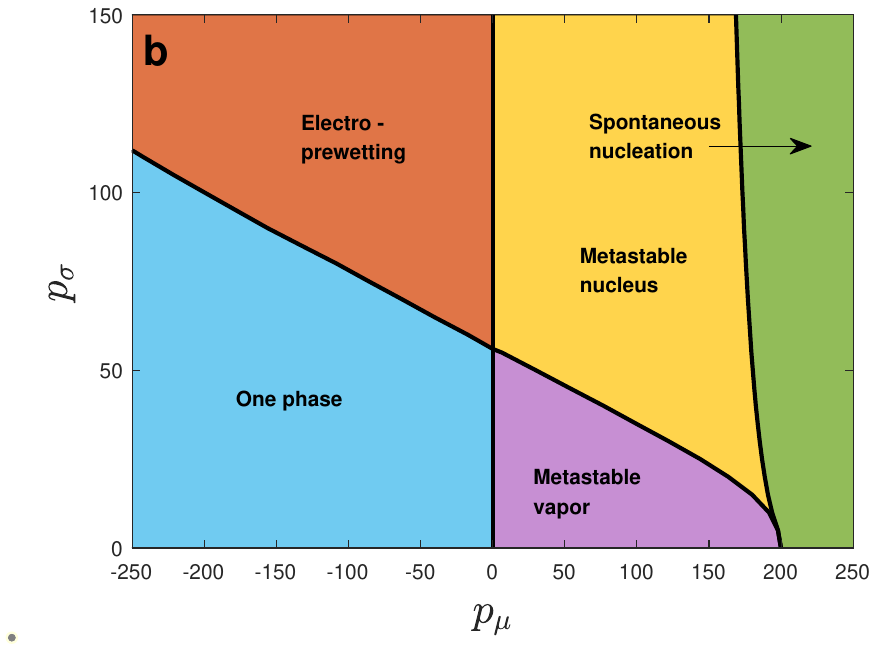}
\caption{State diagram in the $p_{\mu}$--$p_{\sigma}$ (saturation-core charge) plane for $\Delta u=3$ 
(a), and $\Delta u=6$ (b). Note how ionic preferential solvation in (b) enlarges the electro-prewetting 
region and shrinks the one-phase region. For supersaturated fluids, preferential solvation decreases 
the metastable region in favor of metastable nucleus. We used $p_{\gamma}=300$, 
$\lambda_0/R_1=5$, $l_{B0}/R_1=70$, $\veps_{\rm p}=80$ and $\veps_{\rm np}=6$.
}
\label{Phase_diagram}
\end{figure}

Based on the $\Delta G$ curves, one can construct a state diagram in the saturation--colloid charge 
($p_\mu$--$p_\sigma$) plane. Two such state diagrams are shown in Fig. \ref{Phase_diagram} for 
$\Delta u=3$ (a) and $\Delta u=6$ (b). Each region in the state diagrams corresponds to a different  
Gibbs nucleation profile. In the blue region, where $p_{\mu}$ is negative and $p_{\sigma}$ is not large, 
the fluids are in a state of one nonpolar phase. The electro-prewetting zone [orange, Fig. 
\ref{Phase_diagram}(b)] is the region where the fluid is undersaturated while the colloid's charge is 
large enough that field-induced demixing occurs, leading to a finite-sized nucleus. The stability line 
separates these two regions and is defined as $\partial \Delta \tl{G}/\partial \tl{R}|_{\tl{R}=1}=0$. This 
line enters the supersaturated region ($p_{\mu}>0$). At its two sides, the metastable region is divided 
into the metastable vapor region (purple), where $\Delta G(\tl{R})$ does not have a local minimum at 
$\tl{R}>1$, and the metastable polar nucleus region (yellow). 
Spontaneous nucleation occurs at sufficiently large values of $p_\mu$, green region. 
It is separated from the other two regions by the ``electrostatic spinodal'' line. When $p_{\sigma}$ 
tends to zero, the electrostatic spinodal tends to the value of the regular spinodal.

The increase in $\Delta u$ dramatically changes the stability line, enlarging the 
electro-prewetting and metastable nucleus regions. That is, a larger value of $\Delta u$ induces a 
finite-sized nucleus, whether stable or metastable. At the parameter values chosen, the limit between 
super- to undersaturation is not noticeable because, in Eq. (\ref{Full_Gibbs}), $\tl{n}_0(e^{\Delta u}-1)$ 
is of order $0.1$.

\begin{figure}[ht!]
\centering
\includegraphics[width=0.49\textwidth,bb=85 230 500 540,clip]{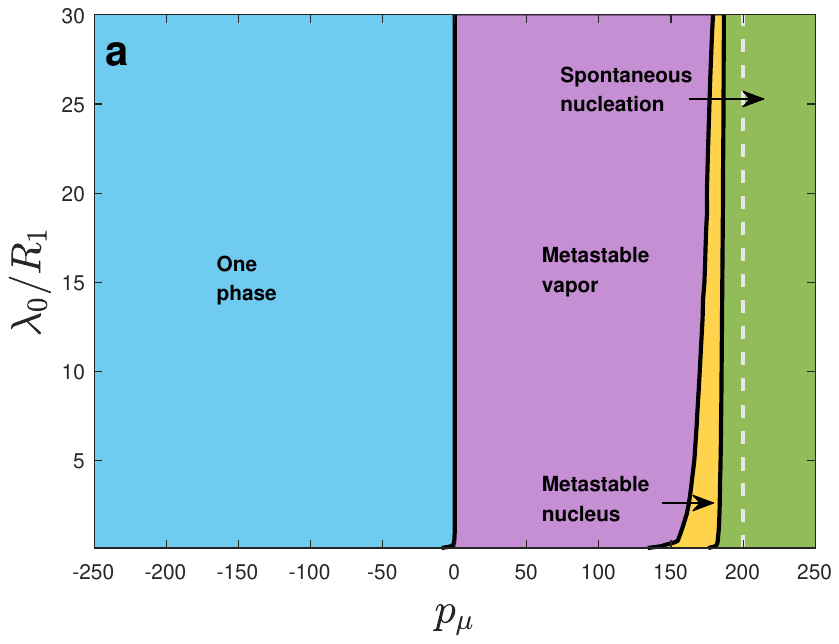}
\includegraphics[width=0.49\textwidth,bb=85 230 500 540,clip]{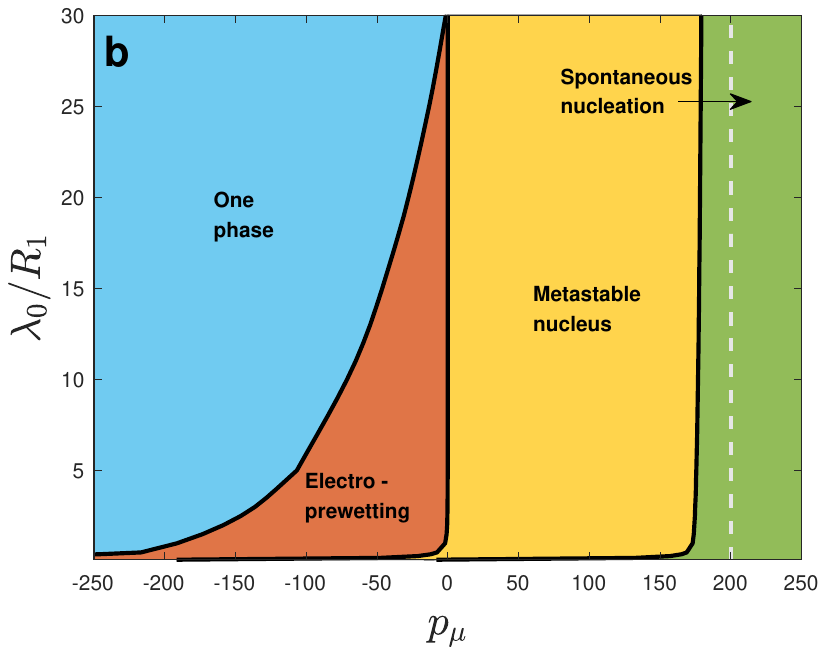}
\caption{State diagram in the $p_{\mu}$--$\tilde{\lambda}_0$ (saturation-Debye length) plane for 
$\Delta u=3$ 
(a), and $\Delta u=6$ (b). The dashed gray line represents the value of the regular spinodal, calculated 
by the classical nucleation theory. We used $p_{\gamma}=300$, $p_{\sigma}=80$, $l_{B0}/R_1=70$, 
$\veps_{\rm p}=80$ and $\veps_{\rm np}=6$.
}
\label{State_diagram_lam}
\end{figure}

We examine the effect of salt concentration by varying the values of $\tl{\lambda}_0$. State diagrams 
in the $p_{\mu}$--$\tl{\lambda}_0$ plane are shown in Figs. \ref{State_diagram_lam}(a) and 
\ref{State_diagram_lam}(b) for $\Delta u=3$ and $\Delta u=6$, respectively, with a fixed value of 
$p_{\sigma}=80$. The stability line shifts significantly due to the dependence on $\Delta u$.
In the absence of a charged core, phase transition occurs with no barrier below the 
regular spinodal line. In the presence of charged particle this line becomes the 
electrostatic spinodal. When this line is crossed by decreasing $p_\mu$, a metastable minimum exists 
with a nucleation barrier. Electro-prewetting is promoted for the stronger solvation preference, 
especially at small $\tl{\lambda}_0$ values (high salt content). The electrostatic spinodal 
is displaced to the left relative to the $\Delta u=3$ case, with a $\approx 12\%$ shift compared to the 
regular spinodal (dashed gray line). 

In the theoretical limit of exceedingly large salt concentrations ($\tl{\lambda}_0\to 0$), spontaneous 
nucleus growth happens for any value of $p_{\mu}$, as the nucleation barrier completely disappears.
Around $\tl{\lambda}_0\approx 3$, the Bjerrum length is equal to the Debye length of the polar phase; 
therefore, for smaller values, the theory should be extended to incorporate ion-ion correlations and 
finite-size effects.\cite{levin_rpp_2002,andelman_jcp_2018,andelman_pre_2019}
Recent works demonstrated increased screening length when a Kirkwood transition occurs.
\cite{rotenberg_2018,Coles_2020,Fetisov_2020}
This increase is similar to an effective increase in $\lambda_0$, and would have little effect in 
Fig. \ref{State_diagram_lam}(a); in Fig. \ref{State_diagram_lam}(b) it would shift the boundary between 
the one-phase and electro-prewetting pahses and would leave in tact the other boundaries.

\subsection{Asymmetry in the Gibbs transfer energy}

We now remove the symmetry assumption in the Gibbs transfer energy, and calculate the Gibbs 
nucleation profiles where $\Delta u^+\neq \Delta u^-$.
In this case, Eq. (\ref{free_energy_int}) can be written as the following dimensionless expression
\begin{eqnarray}
\frac{F_{\rm ions}}{4\pi k_BT}&=&\tilde{n}_0\int\left\{-e^{-\Delta 
u^+(H(\tilde{r}-\tilde{R})-1)-\tilde{\psi}}(\tilde{\psi}+1)\right.\\
&+&\left.e^{-\Delta 
u^-(H(\tilde{r}-\tilde{R})-1)+\tilde{\psi}}(\tilde{\psi}-1)+\veps_i\tilde{\lambda}_0^2(\tilde{\nabla}\tilde{\psi})^2
\right\}\tilde{r}^2d{\tilde{r}},\nn
\label{free_energy_asym}
\end{eqnarray}
where the potential profiles are again found from the modified Poisson-Boltzmann equation, which 
now includes the difference in $\Delta u$.
\begin{figure}[ht!]
\centering
\includegraphics[width=0.49\textwidth,bb=90 230 500 540,clip]{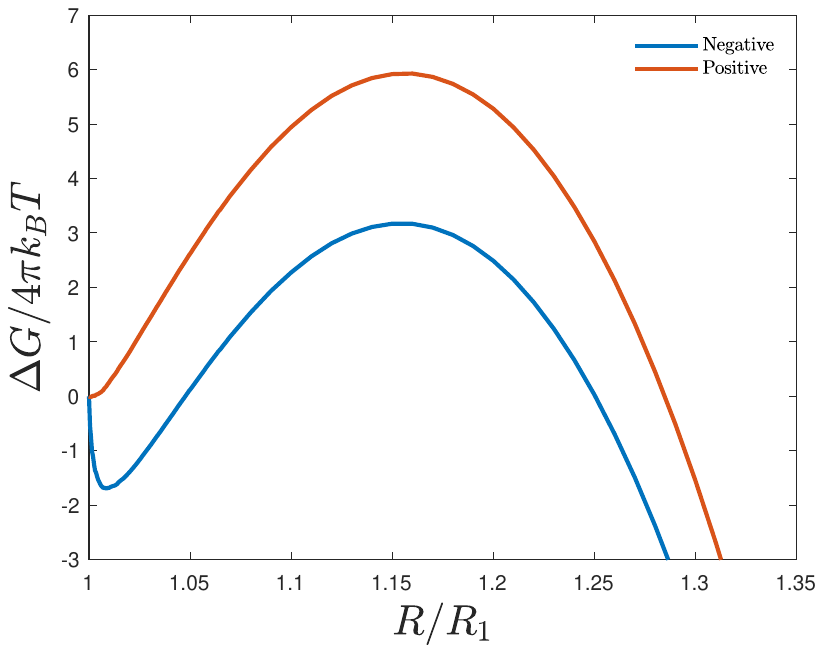}
\caption{Gibbs nucleation energy profiles for a positive and negative core charge sign (red and blue, 
respectively). The Gibbs solvation parameters are $\Delta u^+=6$ and $\Delta u^-=3$, so in both plots, 
the cation's affinity to the polar phase is stronger than the anion's.
We used $p_{\gamma}=350$, 
$\lambda_0/R_1=5$, $l_{B0}/R_1=70$, $\veps_{\rm p}=80$ and $\veps_{\rm np}=6$.
}
\label{Gibbs_sign}
\end{figure}
The sign of the core charge determines whether anions or cations are in excess in the polar phase near 
the core's surface. In the case of asymmetric $\Delta u$s, this means the Gibbs nucleation energy 
curves $\Delta \tl{G}(\tl{R})$ depend on the core's charge.
We calculated $\Delta \tl{G}(\tl{R})$ when $\Delta u^+=6$ for the cations and $\Delta u^-=3$ for the 
anions. We used both a positive and negative sign for the core particle surface charge $\sigma$.
The profiles are shown in Fig. \ref{Gibbs_sign}. For the parameter values chosen, the energy barrier for 
the positive core 
charge is $\Delta G^*/k_BT\approx5.9$ while for the negative one it is $\Delta G^*/k_BT\approx4.9$. 
The nucleation rate almost triples due to the exponential dependence on the nucleation barrier.

\section{Summary and conclusions}

We propose a simple mesoscopic model to describe nucleation induced by a charged core in polar 
fluids. The model applies to liquid/liquid coexistence and to a lesser extent to liquid/vapor systems.
The solid spherical core is uniformly charged and surrounded by two layers of fluids separated by a 
thin interface, as is appropriate for a system far from its critical point. The fluids' properties are 
characterized by their bulk values. The main novelty of the present work is the inclusion of the ions' 
preferential solvation and electrostatic energy into the nucleation energy. The electrostatic potential 
around the colloid obeys a modified Poisson-Boltzmann equation. The equation is solved analytically 
for small charges and numerically otherwise. Once the potential is known, the ionic contribution is 
calculated and added to the interfacial and bulk energies to yield the total Gibbs free energy of 
nucleation. 

The Thomson model for purely dielectric fluids shows that the energy barrier for nucleation decreases 
in the presence of a core-charged particle. As we show, in polar fluids, the solvation energy of ions 
amplifies this tendency despite the screening of the core charge. The colloid potential {\it decreases} 
monotonically with $\Delta u$. The state diagram in the plane of saturation and core charge is found 
from the Gibbs energy profiles. We show that preferential solvation changes considerably the lines 
separating the various regions--one phase, electro-prewetting, metastable vapor, metastable nucleus, 
and spontaneous nucleations, see Fig. \ref{Phase_diagram}. In the phase diagram, the 
electrostatic spinodal separates the regions of metstable nucleus and the spontaneous nucleation. 
Upon crossing it, the energy barrier for nucleation disappears. In the absence of charge, this line 
becomes identical with the classical spinodal in bulk systems.
In addition, differences in the preferential solvations of cations and anions, incorporated in the model 
using specific values of $\Delta u^+$ and 
$\Delta u^-$, may explain the asymmetry in nucleation rates between positive and negative cores.
\cite{kenjiro_gra_2006,fernandez_jcp_2002}

Our continuum model uses point-like ions and predicts nucleus-charge sign asymmetry, 
and this is not found in experiment with sub-$3$nm particles.\cite{Kangasluoma_2016} This can be due 
to the finite-ions size and other effects. For example, the solvation energy and permittivity can vary on 
the nanometer scales, and ion complexation with other ions or solvent molecules is an intricate 
ion-specific process. Small clusters of ions and solvent molecules can act as the charged particle used 
in our model, though their shape will deviate from a perfect sphere. A possible multi-scale approach 
for future studies could be the integration of molecular dynamics or density functional theory 
simulations at these small scales with our continuum model to provide a comprehensive view of the 
nucleation process. 

The current model assumes that the nucleating particles are spherical. Deviations from 
spherical symmetry could significantly modify the nucleation energies and rates and should be 
studied. A similar effect could occur with geometrically spherical particles with inhomogeneous surface 
charge distribution.

The expression for nucleation in the presence of ions Eq. (\ref{Full_Gibbs}) is our main analytical result. 
The preferential solvation of ions reduces the energy barrier, assisting nucleation. 
Our previous work has shown that when a sufficiently large quantity of ions is present, without 
preferential solvation, screening of the core particle's electric charge results in a behavior similar to 
that of classical nucleation theory, only with a renormalized interfacial tension between the polar fluid 
and the particle's surface.\cite{KROLL_2023} A similar effect occurs when $\Delta u>0$, where 
$p_\sigma h(\tl{R})$ reduces to a surface effect, see Eq. (\ref{eff_ST}). 

Preferential ion solvation appears explicitly in the volume term $\sim (\tl{R}^3-1)$ in Eq. 
(\ref{Full_Gibbs}) as a term that renormalizes the value of $p_\mu$, i.e., ions shift the location of the 
binodal curve. This shift is equivalent to a change of the chemical potential, Eq. (\ref{eff_mu}). This shift 
can be quite large due to the exponential dependence $\sim \exp(\Delta u)$, recalling that in some 
liquids and ions $\Delta u$ can be of order $10$.\cite{Marcus_1989}

When the Bjerrum length is large, for example, when the nonpolar phase is vapor, the model should 
include ion correlations. In addition, 
the dependence of the Gibbs transfer energies and the liquid's dielectric constants on ion density 
should be considered where relevant.\cite{andelman_jcp_2011,andelman_jcp_2015b} 
For systems close to the critical point, the electric field gradients lead to notable deviations 
from the coexistence line even in the absence of ions 
\cite{landau_book,tsori_nature_2004,tsori_jcp_2022} and the resulting first- and second-order phase 
transitions \cite{tsori_jcp_2021,tsori_pre_2021} should be described using the density functional 
theory.\cite{wu_2008,tarazona_mol_phys_1984,oxtoby_jcp_1994,evans_jcp_2004,onuki_jpcm_2016}

It is reasonable to assume that a salt with a large affinity $\Delta u$ can accelerate nucleation not only 
by reducing the energy barrier of a single nucleus but also through the coalescence of 
numerous metastable nuclei. At large distances, these nuclei repel because they are 
similarly-charged. 
At small distance, however, coalescence could occur if capillary forces are sufficiently strong.
It can be interesting to look at the coalescence of such droplets in multi-particle 
theory or simulation. Moreover, it is intriguing whether such coalescence could occur in the 
electro-prewetting region of the state diagram.
Antagonistic ions, where the cation prefers one phase, 
and the anion prefers the other, can have consequences to nucleation. For such ions, the anions and 
cations partition across the interface, reducing the surface tension between liquids, and this may 
further increase the nucleation rates.\cite{Kroll_2020,kroll_epje_2023}

\section*{Acknowledgments}

We are grateful for financial support by the Israel Science Foundation via Grant No. 332/24.

\bibliographystyle{unsrt} 
\bibliography{Nucleation_solvation2}

\end{document}